\documentclass[prc,nofootinbib]{revtex4}

\usepackage{epsfig,amsmath,color,multirow}

\def\snn{$\sqrt s_{\rm NN}$}
\def\epem{${\rm e}^+ {\rm e}^-$}

\newcommand{\be}{\begin{equation}}
\newcommand{\ee}{\end{equation}}                                                                               
\newcommand{\bea}{\begin{eqnarray}}
\newcommand{\eea}{\end{eqnarray}} 

\begin{document}

\title{Centrality dependence of hadronization and chemical freeze-out 
conditions \\
in heavy ion collisions at $\sqrt s_{NN} = 2.76$ TeV.} 

\author{Francesco Becattini}
\affiliation{Universit\`a di Firenze and INFN Sezione di Firenze, Firenze, Italy}

\author{Marcus Bleicher}
\affiliation{Frankfurt Institute for Advanced Studies (FIAS), Frankfurt, Germany}

\author{Eduardo Grossi}
\affiliation{Universit\`a di Firenze and INFN Sezione di Firenze, Firenze, Italy}

\author{Jan Steinheimer}
\affiliation{Frankfurt Institute for Advanced Studies (FIAS), Frankfurt, Germany}

\author{Reinhard Stock}
\affiliation{Frankfurt Institute for Advanced Studies (FIAS), Frankfurt, Germany}
\affiliation{Institut f\"ur Kernphysik, Goethe-Universit\"at, Frankfurt, Germany}

\begin{abstract}
We present an analysis of hadronic multiplicities measured in Pb-Pb collisions
at $\sqrt s_{NN} = 2.76$ TeV as a function of the collision centrality within the 
statistical hadronization model. Evidence is found of a dependence of the chemical
freeze-out temperature as a function of centrality, with a slow rise from central
to peripheral collisions, which we interpret as an effect of post-hadronization 
inelastic scatterings. Using correction factors calculated by means of a simulation
based on the UrQMD model, we are able to obtain a significant improvement in the 
statitical model fit quality and to reconstruct the primordial chemical equilibrium 
configuration. This is characterized by a nearly constant temperature of about 164 MeV 
which we interpret as the actual hadronization temperature.
\end{abstract}

\maketitle

\section{Introduction}

The determination of the critical temperature of QCD is one of the principal goals 
of relativistic nucleus-nucleus (AA) collision physics. This temperature has been 
calculated in lattice QCD \cite{karsch,fodor,delia} to be about 160 MeV, at a 
baryon-chemical potential $\mu_B \simeq 0$, a situation characteristic of heavy ion 
collisions at top RHIC and LHC energies. At very low $\mu_B$ the phase transition 
from hadronic matter to Quark-Gluon Plasma(QGP) has been found to be a continuous 
one (a {\it cross-over}), so that the value of the (pseudo-) critical temperature 
depends somewhat on the specific observable under consideration \cite{karsch,fodor}. 

It has been conjectured for quite some time \cite{stock} that the experimentally measured 
hadronic multiplicities, or multiplicity ratios do, in fact, represent such an 
observable: they depend on the temperature prevailing at, or near QCD hadronization. 
It has been proposed to use fluctuation of conserved charges to determine it 
\cite{karschgroup,ratti} as these can be directly calculated in lattice QCD. However, 
multiplicities are first moments and, as such, are more robust observables against 
spurious effects. 
Indeed, a statistical {\it ansatz} is able to reproduce the measured hadronic yields, both 
in elementary \cite{becaelem} and in relativistic nucleus-nucleus collisions 
\cite{various}. This has led to the formulation of the Statistical Hadronization 
Model(SHM) which, in a nutshell, assumes that hadrons are emitted from the fireball 
source at (almost) full chemical equilibrium. The reason of such a success, unexpected
in elementary collisions, as well as the identity of the fitted temperature in all 
kinds of collisions, has been debated for a long time (see refs.~\cite{satz,becareview} 
for a summary). In practice, one can take advantage of this phenomenon to obtain 
the position of the parton-hadron coexistence line of QCD matter in the $(T,\mu_B)$ 
plane.

The temperature determined by fitting the hadronic multiplicities with the SHM is 
actually the one at which hadrons/resonances cease inelastic interaction, the so-called 
"chemical freeze-out" temperature. In principle this may differ from the QCD transition 
temperature if hadrons, after their formation, keep interacting inelastically. This 
is, clearly, not the case in elementary \epem annihilation to hadrons but it could 
become relevant in the high multiplicity final state of AA collisions. Different 
reactions could then freeze-out at different times, in inverse order of inelastic 
cross section, so that this stage of the fireball source expansion, dubbed as "afterburning", 
would generally imply deviations from full chemical equilibrium of the hadronic 
species \cite{bass}. In the standard SHM analysis such effects were assumed to be 
negligibly small, and that, therefore, the temperature and baryon-chemical potential
yielded an ideal snapshot of the fireball dynamical trajectory, at or near QCD 
hadronization.

An unexpected recent outcome from LHC has been the relatively low ${\rm p}/\pi$ ratio 
measured by the ALICE experiment in central Pb+Pb collisions \cite{milano} at \snn=2.76 
TeV, with respect to the expectation from the SHM \cite{pbmpred}. A similar result was 
obtained earlier by the SPS experiment NA49 \cite{na49pbar} which reported sizeably 
low $\bar{\rm p}$ and ${\bar \Lambda}$ yields compared to SHM predictions \cite{becapred}. 
This has been interpreted \cite{becafrank1,steinheimer,stockproc,pratt} as an evidence 
of post-hadronization baryon-antibaryon annihilation. An alternative explanation has been
put forward in ref.~\cite{rafelski}. Note 
that annihilation cross sections do not fade away with dropping temperature, unlike 
inelastic transmutations. In ref.~\cite{becafrank2} we proposed a picture of hadron 
production in relativistic A+A collisions based on the idea of the hadronization 
process leading to chemical equilibrium of its outcome, followed by a stage of afterburning 
driving some hadronic species (notably baryons and antibaryons) out of chemical 
equilibrium before freeze-out. We determined these effects by employing the hybrid 
version of the microscopic transport model UrQMD \cite{urqmd}, obtaining {\em modification
factors} due to afterburning which were then employed in the subsequent data analysis. 
We thus reconstructed the primordial chemical equilibrium, up to a point where 
multi-hadron collisions could become important, and showed, for central collisions at 
various energies, a resulting rise of the deduced temperatures, and significantly 
improved SHM fit quality. This bears out the idea of a primordial chemical equilibrium 
as an intrinsic feature of hadronization.

In the present paper we extend our analysis, changing the topic from central collisions 
at various energies, to consideration of the centrality dependence of hadron multiplicities 
at fixed energy. Whereas, in central collisions, the final hadronic expansion stage 
causes substantial antibaryon and (at higher energies) baryon annihilation/regeneration,
these effects should diminish toward more peripheral collisions because of the reduced
overall multiplicity (see discussion in sect.~\ref{fo}). Thus, if our hypothesis is correct that 
the QCD hadronization process generates an equilibrium hadron/resonance yield distribution, 
at some constant temperature $T$, the afterburning effects should lead to a larger  
modification in central than in peripheral collisions. As baryon attenuation leads 
to lower apparent freeze-out temperatures derived from the standard SHM analysis, we 
would expect this temperature to rise, mildly, from central toward peripheral collisions. 

These expectations have been borne out by a detailed hydrodynamical investigation 
\cite{heinzkestin} of the final stages of AA collisions. Interestingly, it was pointed out 
in ref.~\cite{heinzkestin} that, if afterburning played some role, the chemical 
freeze-out temperature $T_{\rm chem}$ fitted within the SHM should exhibit a non-trivial 
behaviour, with a rise from central to peripheral collisions. Indeed, this effect 
is clearly observed for the {\em kinetic} freeze-out temperature, the temperature at which 
hadrons cease their elastic interactions. On the other hand, at the highest RHIC 
energy, no significant dependence of $T_{\rm chem}$ on centrality was seen 
\cite{heinzkestin,starfits,becamann}, indicating that chemical composition are much 
less affected than spectra by the afterburning stage. In fact, the STAR experiment has 
found a dependence of $T_{\rm chem}$ on centrality at lower energy \cite{bes}, but 
the slope of the function is reversed if the strangeness neutrality is enforced. 
It should also be kept in mind that at low energy the use of midrapidity densities 
may give rise to spurious effects such as an artificial enhancement of the strange 
particles, so that this observed dependence is difficult to intepret at this time.

Recently, the ALICE experiment at the LHC has provided \cite{aliceaa} a set of 
high precision measurements of hadronic species midrapidity multiplicities, as a 
fuction of centrality in Pb+Pb collisions at \snn = 2.76 TeV. The improved accuracy
and the increased total multiplicity with respect to RHIC energy, should allow to
highlight a dependence of $T_{\rm chem}$ on centrality. It is precisely the goal 
of this paper to test the centrality dependence of the chemical freeze-out temperature. 
This should settle the much debated ``proton anomalies", and provide further evidence 
for the constancy of the primordial hadronization temperature, to be identified 
with the pseudo-critical QCD temperature. To this end we first analyze the ALICE 
data with the standard SHM method. Then, by employing modification factors for all 
hadronic species, and all centralities, obtained from the hadronic transport model 
UrQMD, we shall show that significant modification of the primordial abundances occurs 
in central collisions, in agreement with the findings in refs.~\cite{becafrank1,becafrank2}, 
reducing towards more peripheral collisions and mostly affecting the baryon-antibaryon 
species, via annihilation and regeneration. With the modification factors in place 
in a second SHM analysis we shall arrive at a uniform temperature of 164$\pm$3 MeV.

\section{The freeze-out process\label{fo}}

We can understand the effect of multiplicity on chemical freeze-out in relativistic
heavy ion collisions with simple arguments.
In an expanding sytem of interacting particles freeze-out occurs when the mean scattering 
time $\tau_{\rm scatt}$ exceeds the mean collision time $\tau_{\rm exp}$:
\be
   \tau_{\rm scatt} = \frac{1}{n \sigma \langle v \rangle} > \tau_{\rm exp}
    = \frac{1}{\partial \cdot u}
\ee
$u$ being the hydrodynamical velocity field and $\langle v \rangle$ is the mean 
velocity of particles. If the cross-section $\sigma$ is the inelastic one, the 
freeze-out is called {\em chemical}, whereas if it includes elastic processes, the 
freeze-out is called {\em kinetic}. Chemical freeze-out of course precedes the kinetic
as the inelastic cross section is smaller than the total. 

We can obtain a gross approximation of the expansion time with the ratio $V/\dot V$
where $V(t)$ is the volume of the fireball at the time $t$. For a fireball which 
is spherical in shape with a radius $R$, this is $R/3\dot R$ and if the radius increases
at approximately the mean particle velocity $\langle v \rangle$, we have the condition:
\be\label{freeze}
   \frac{1}{n \sigma \langle v \rangle} > \frac{R}{3 \langle v \rangle}
   \implies \frac{1}{n \sigma} > \frac{R}{3}
\ee
For a given number of particles $N$ within the volume, this inequality yields the
radius at which freeze-out occurs as a function of $N$ and of the average cross-section:
\be\label{rfo}
     R_{\rm fo} = \sqrt{\frac{N \sigma}{4 \pi}}
\ee
and the density at which freeze-out occurs, which decreases with $N$ according to:
\be\label{fodens}
  n_{\rm fo} = \frac{N}{\frac{4\pi}{3} R^3_{\rm fo}} = 3 \sqrt{\frac{4\pi}{N}} 
  \frac{1}{\sigma^{3/2}}
\ee  
Of course, it should be kept in mind these estimates (\ref{rfo}) and (\ref{fodens}) 
are crude, but they tell us that the freeze-out radius, for each particle, approximately 
scales with the square root of the number of scattering centers a particle can interact 
with and the related cross section. For a low multiplicity hadronic system, it may happen 
that the above value exceeds the density of hadrons when they are formed, that is at hadronization.
This simply signals that hadrons decouple right after their formation without reinteracting,
what happens in elementary collisions, at the intrinsic hadronization density scale 
which is dictated by QCD. For relativistic heavy ion collisions, conversely, the 
multiplicity can grow to large numbers so that there could be enough time for hadronic
reinteraction and freeze-out occurs later. For instance, for the typical value of $N=1000$, 
in most central collisions, and $\sigma= 30 \, {\rm mb} = 3 \, {\rm fm}^2$ 
one has $R_{\rm fo} \simeq 15$ fm, which is in the right ballpark (for kinetic freeze-out)
taking into account the drastic approximations made; the density at freeze-out turns 
out to be $n_{\rm fo} \simeq 0.06$ fm$^{-3}$ which is lower than the typical hadronization 
density of about 0.5 fm$^{-3}$. 
\begin{figure}[ht]
\begin{center}
\includegraphics[width=0.6\columnwidth]{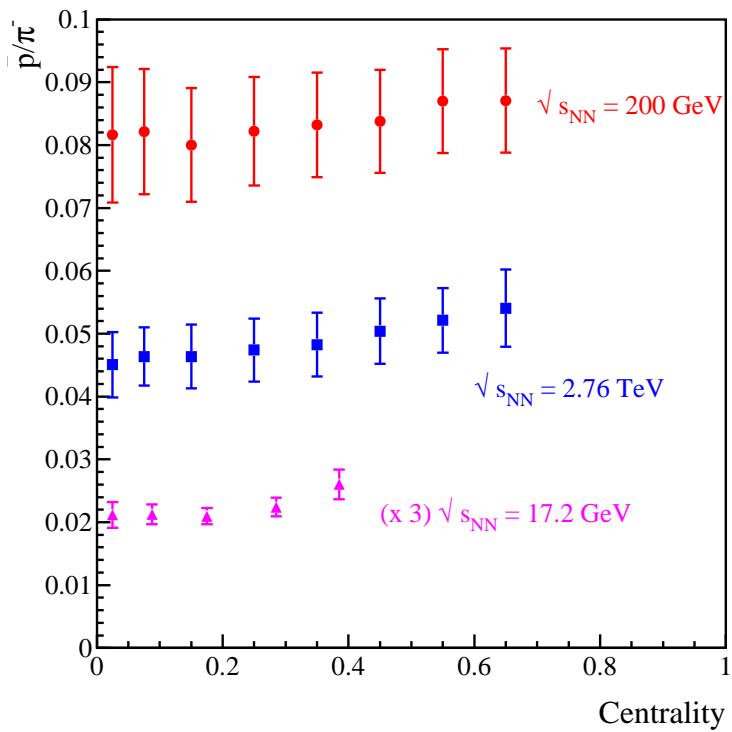}
\caption{(Color online) Ratio between antiproton and negative pion yields in 
relativistic heavy ion collisions as a function of centrality at different energies.}
\label{pbarp}
\end{center}
\end{figure} 

The above equations also imply that, if hadronization occurs at a universal temperature 
$T_h$ \cite{becareview,satz} which is the pseudo-critical QCD temperature, the effective 
average temperature of the chemical freeze-out should increase in peripheral collisions 
if equilibrium is approximately
maintained in the hadronic reinteraction stage. The strength of this effect depends, 
according to (\ref{fodens}), on the function $n_{\rm fo}(T)$ and it is, as expected, 
larger for the kinetic than chemical freeze-out simply because the total cross section 
is larger than the inelastic one. In general, since the hadronic density strongly 
depends on the temperature, the dependence of $T_{chem}$ on $N$, hence on centrality, 
is mild. To highlight it, one needs a large lever arm in terms of multiplicity and 
higher energies are more favourable in this respect, as has been mentioned in the Introduction.
\begin{table}[!ht]\begin{center}
\begin{tabular}{c|c|c|c|c}
\hline\hline
 \snn (GeV) & $\bar{\rm p}/\pi^-$ (AA) & $\bar{\rm p}/\pi^-$ (pp) & $\bar \Xi^+/\pi^-$ (AA) & $\bar \Xi^+/\pi^-$ (pp) \\ 
\hline
 17.2         &  $0.0067\pm 0.00062$ \cite{na49aa} &  $0.0165\pm 0.0005$ \cite{na49pp} & $(1.12 \pm 0.17) 10^{-3}$ \cite{na49aa} & $(3.9 \pm 0.4) 10^{-4}$ \cite{na49pp}  \\
 200          &  $0.082 \pm 0.012$ \cite{staraa} &  $0.080 \pm 0.009$  \cite{starpp} &  $(6.6 \pm 0.79) 10^{-3}$ \cite{staraa} &  $(2.0 \pm 0.7) 10^{-3}$ \cite{starpp} \\
 2750         & $0.045 \pm 0.005$ \cite{aliceaa} & - & $(4.7 \pm 0.5) 10^{-3}$ \cite{aliceaa} &  -  \\
 7000         &  -  & -  &  - & $(3.25 \pm 0.32) 10^{-3}$ \cite{alicepp} \\
\hline\hline
\end{tabular}\end{center}
\caption{Ratios $\bar{\rm p}/\pi^-$ and $\bar \Xi^+/\pi^-$ in pp and AA collisions
at different energies. The ratio $\bar \Xi^+/\pi^-$ in pp is always less than in
AA at the same \snn.}
\label{pbarpi}
\end{table}
\begin{figure}[ht]
\begin{center}
\includegraphics[width=0.6\columnwidth]{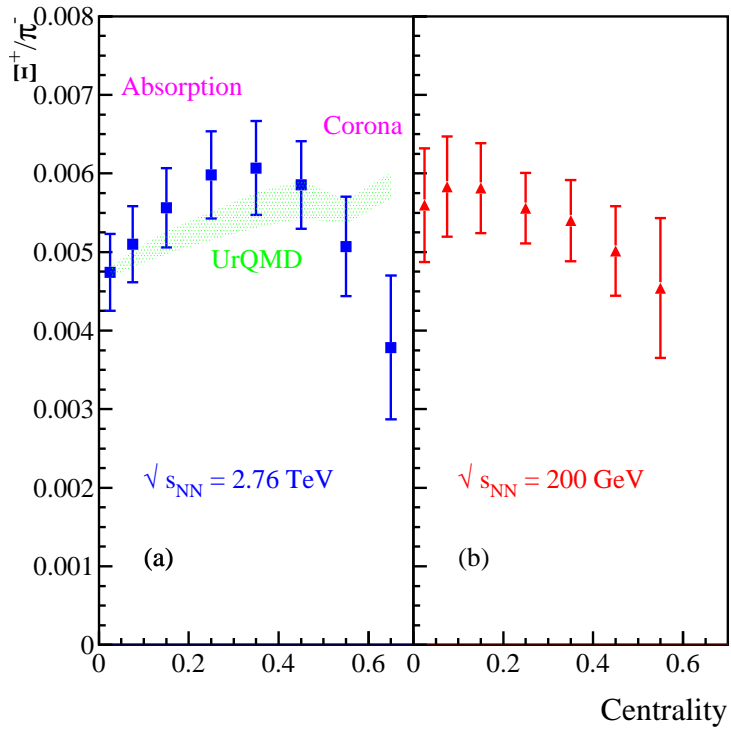}
\caption{(Color online) Ratio between $\bar\Xi^+$ and negative pion yields in relativistic heavy 
ion collisions as a function of centrality at different energies. At the largest
energy (a), a structure can be clearly seen. We interpret the bump in mid-peripheral
events as the combination of two effects: increased baryon annihilation owing to
larger multiplicity at LHC and the corona effect in very peripheral. Also shown
the prediction of corona-less UrQMD calculations normalized to the most central bin.}
\label{xip}
\end{center}
\end{figure} 

Before moving to the data analysis, it should be pointed out that there is evidence of
afterburning in the data itself. In fig.~\ref{pbarp} we show the behaviour of 
${\rm \bar p}/\pi^-$ ratio as a function of centrality at different centre-of-mass 
energies. In all three cases, the ratio slightly, yet significantly (taking into account 
that errors are visibly correlated) increases from central to peripheral collisions, 
in agreement with the expectation of a larger anti-baryon annihilation in more central 
events. Note that, at the LHC energy, this effect cannot by any means be explained 
by a genuine decrease of baryon-chemical potential at freeze-out in peripheral collisions 
because all particle/antiparticle ratios are consistent with $\mu_B \simeq 0$ at all 
centralities. 
A possible mundane explanation to be considered is a core-corona superposition 
if the ratio $\bar{\rm p}/\pi^-$ was larger in pp than central AA at the same energy 
(see table~\ref{pbarp}). However, this is ruled out by the stunning centrality behaviour 
of the ratio $\bar \Xi^+/\pi^-$, shown in fig.~\ref{xip}. Unlike at RHIC, this ratio 
surprisingly {\em increases} from central towards peripheral collisions, then drops 
according to the expectations of the core-corona model \cite{becacorona} as its value 
is indeed much lower in pp than in AA collisions (see table~\ref{pbarp}) at all energies.

The rise of the $\Xi/\pi$ ratio is the result of the larger relative absorption of 
$\Xi$ and the larger relative production of pions in the most central collisions. 
Altogether, the centrality dependence of these particle ratios confirm the expected 
dependence of chemical freeze-out on particle multiplicity.

\section{Data analysis}
\label{analysis}

We have analyzed the multiplicities measured by the ALICE experiment at \snn = 2.76
TeV \cite{aliceaa} to determine the chemical freeze-out parameters with fits
to the usual SHM (described in ref.~\cite{becareview}) predictions and to the same 
formulae corrected for the modification factors, defined as the ratios between the 
particle yields with afterburning and the same yields without it. The modification factors 
have been estimated with a hybrid version of the code UrQMD \cite{urqmd} implementing
afterburning after a hadron generation according to local thermodynamical equilibrium
prescription (Cooper-Frye formula). Therefore, the estimated factors are the outcome
of a full simulation of the heavy ion collision process  

\subsection{Data interpolation}

The midrapidity densities of hyperons \cite{aliceaa} are provided by the ALICE experiment 
with a centrality binning different from that of p, K and $\pi$ (10 centrality 
classes for the latter, 7 for $\Lambda$'s and 5 for $\Omega$ and $\Xi$'s). Thus, we 
have interpolated the yields of hyperons to obtain their values in the same centrality 
bins as for the protons, pions and kaons. As interpolation function we chose a 6th 
degree polynomial for $\Lambda$'s and a 4th degree polynomial for $\Omega$'s and $\Xi$'s. 
In order to make a proper comparison with the data, we calculated the integral mean 
value within each bin:
$$
N([c_i,c_{i+1}])=a_0+a_1(c^2_{i+1}-c^2_{i})/2(c_{i+1}-c_{i})+
a_2(c^3_{i+1}-c^3_{i})/3(c_{i+1}-c_{i})+\cdots
$$
where $c_i$ are the centrality limits of each bin. We have determined the coefficients $a_i$
by making a $\chi^2$ fit to the data in the various centrality bins. Since the experimental 
errors among different centrality bins are apparently correlated, we have formed a 
non-diagonal covariance matrix $C$ in the $\chi^2$:
$$
\chi^2 = \sum_{\rm bins} ({\rm Theo_i} - {\rm Meas}_i) C^{-1}_{ij} 
({\rm Theo_j} - {\rm Meas}_j)
$$
assuming a constant correlation coefficient $\rho = 0.5$. Once the parameters $a_i$ 
of the interpolating function were obtained, we have been able to estimate the yields 
of the hyperons along with their error in the same bins of proton, pions and kaons 
(see table \ref{tab:int}) up to the 70-80\% bin. Since the data from ref.~\cite{aliceaa} 
show that the yields of particle and anti-particles are compatible within errors, we
have interpolated the sum $\Omega+\bar{\Omega}$ to reduce the statistical uncertainty
in the interpolation.
\begin{table}
\setlength{\tabcolsep}{10pt}
\centering
\begin{tabular}{l|c|c|c|c}
\hline \hline
& $\Lambda$  & $\Xi^-$  & $\bar{\Xi}^+$  & $\Omega+\bar{\Omega}$ \\
\hline
0-5\%& 
 26.1 $\pm$ 2.8 &3.57 $\pm$ 0.27  &  3.47 $\pm$ 0.26 & 1.26 $\pm$ 0.22  \\
		 5-10\%& 
 22.0 $\pm$ 1.9 & 3.13 $\pm$ 0.21 &  3.08 $\pm$ 0.20 & 1.07 $\pm$ 0.17  \\
		 10-20\%&
 17.1 $\pm$ 1.6 & 2.52 $\pm$ 0.15 &  2.52 $\pm$ 0.15 & 0.83 $\pm$ 0.12  \\
		 20-30\%&
 12.0 $\pm$ 1.1 & 1.80 $\pm$ 0.12 &  1.83 $\pm$ 0.12 & 0.57 $\pm$ 0.09  \\
		 30-40\%&
 8.0 $\pm$ 1.0 &  1.19 $\pm$ 0.09 &  1.22 $\pm$ 0.09 & 0.37 $\pm$ 0.07  \\
		 40-50\%&
 4.9 $\pm$ 0.5 &  0.70 $\pm$ 0.05 &  0.72 $\pm$ 0.05 & 0.22 $\pm$ 0.04  \\
		 50-60\%&
 2.7 $\pm$ 0.5 &  0.35 $\pm$ 0.04 &  0.36 $\pm$ 0.04 & 0.120 $\pm$ 0.031  \\
60-70\%&
1.32 $\pm$ 0.35 & 0.149 $\pm$ 0.034 &  0.140 $\pm$ 0.033 & 0.053 $\pm$ 0.025  \\
		 70-80\%&
0.68 $\pm$ 0.35 & 0.099 $\pm$ 0.034 &  0.100 $\pm$ 0.034 & 0.011 $\pm$ 0.025  \\
\hline\hline
\end{tabular}
\caption{Results of the interpolation for the midrapidity yields of $\Omega$, 
$\Xi$, $\Lambda$ in the same centrality class of proton, kaon and pion 
\cite{aliceaa}. }\label{tab:int}
\end{table} 
\begin{figure}[ht]
\begin{center}
\includegraphics[width=0.6\columnwidth]{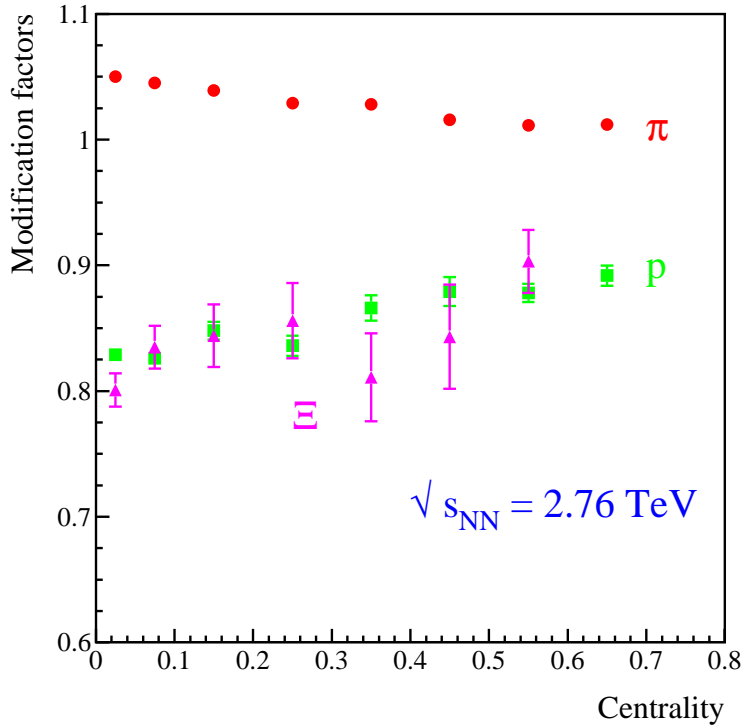}
\caption{(Color online) Modification factors (see text for definition) for $\pi^+$, proton, and $\Xi^-$ 
as a function of centrality at \snn = 2.76 TeV calculated with UrQMD. The error bars are
statistical. }
\label{urqmfact}
\end{center}
\end{figure} 

\subsection{Calculation of modification factors}

To quantify the effects of the hadronic phase (afterburning) on the particle ratios 
we employ the Ultra relativistic Quantum Molecular Dynamics model (UrQMD) in its current 
version \cite{urqmd}. The hadronic transport part of the model is based on an effective 
solution of the relativistic Boltzmann equation
\be\label{boltzmann}
   p^\mu \partial_\mu f_i(x^\nu, p^\nu) = \mathcal{C}_i \quad .
\ee
which describes the time evolution of the distribution functions $f_i(x^\nu, p^\nu)$ 
for particle species $i$, including the full collision term on the right hand side. 
The interactions of hadrons in the current version is limited to binary elastic 
and $2 \rightarrow n$ inelastic scatterings, including resonance creations and 
decays, string excitations and particle-antiparticle annihilations. 
The cross sections and branching ratios for the corresponding interactions are taken 
from experimental measurements, where available, and detailed balance relations.\\ 
The modification factors, required for our analysis, are extracted by running the 
fluid dynamics mode of the UrQMD hybrid model, as discussed in ref.~\cite{steinheimer}, 
for Pb+Pb collisions at $\sqrt{s_{NN}}= 2.76$ TeV and the centralities defined by 
the ALICE experiment. 

We then analyze the particle multiplicities of stable hadrons 
either at the end of the fluid dynamical phase, or after the hadronic rescattering 
phase of the nuclear collision. The transition point from the fluid dynamical phase 
to the hadronic transport part occur in successive transverse slices, of thickness 
0.2 fm, whenever all fluid cells of that slice fall below a critical energy density, 
that is six times the nuclear ground state density $\epsilon \approx 850 $ 
$\rm{MeV/fm^3}$ (in accordance with measures particle yields \cite{steinheimer}),
which is then the maximal energy density at which particles are generated. For the 
hydrodynamical stage of the UrQMD simulation we have applied an EoS that follows 
from combining a hadronic phase with an effective mean field quark model, see 
ref.~\cite{steinh2}. In the UrQMD hybrid model hadrons of species $i$ are produced 
by sampling the particle distributions defined by the Cooper-Frye prescription on 
a pre-defined hypersurface $\sigma_{\mu}$: 
\be\label{cooper}
        E \ \frac{dN}{d^3p} = g_i \int_{\sigma}{f_i(x,p) \ p^{\mu} \ d\sigma_{\mu}}
\ee
Serving as an input, the local temperature, the chemical potentials and the flow 
velocity $u_{\mu}$ enter the particle distribution function $f_i$, i.e. all the particles, 
at the end of the fluid dynamical phase, are produced according to local chemical 
equilibrium. We therefore obtain the particle yield $N_i$ either at the latest chemical
equilibrium point (LCEP, see ref.~\cite{becafrank2})$N_i^{CE}$ or after the chemical 
and kinetic freeze out $N_i^{FO}$. The modification factor $F_i^s$, of particle species 
$i$ is then simply defined as $F_i^s=N_i^{FO}/N_i^{CE}$. Note that the modification 
factors have been determined by turning off weak decays but performing all strong 
decays, in accord with the yields quoted by the ALICE experiment. It should 
also be stressed that in this procedure the UrQMD average midrapidity particle 
multiplicities, generated at the end of the hydrodynamical stage (after the Cooper-Frye 
procedure) do, indeed, exhibit a common temperature of 158.2$\pm$2.2MeV if we fit them 
with the statistical model, as shown in table~\ref{check}. Therefore, our calculated 
modification factors are close to those which would result from a calculation at the 
actually determined latest chemical equilibrium temperature of about 164 MeV 
in the data analysis (see sect.~\ref{results})\footnote{We note in passing that the 
average Cooper-Frye temperature in ref.~\cite{steinheimer} was obtained by calculating 
an average of the temperatures in the various hydro cells weighted with pion yields and it 
is therefore not directly comparable with the temperature determined by fitting 
particle multiplicities.}

\begin{table}[!ht]\begin{center}
\begin{tabular}{|c|c|c|}
\hline
   Temperature               &   $\mu_B$         &   $\chi^2$/dof     \\
\hline
    158.2 $\pm$ 2.2  MeV     &     0 (fixed)     &   2.52/8    \\
\hline\hline
   Particle                  &   Calculated      & Fitted with SHM  \\
\hline
   $\pi^{+}$                 & 528$\pm$37          & 542.4   \\
   $\pi^{-}$                 & 529$\pm$37          & 542.4    \\
   K$^{+}$                   & 100.0$\pm$7.7       & 95.63    \\
   K$^{-}$                   & 101.0$\pm$7.7       & 95.63    \\
   p                         & 33.7$\pm$2.4        & 33.31     \\
   $\bar{\rm p}$             & 30.9$\pm$2.2        & 33.31     \\
   $\Lambda$                 & 18.9$\pm$1.6        & 18.45     \\
   $\Xi^{-}$                 & 2.79$\pm$0.19       & 2.744     \\
   $\Xi^{+} $                & 2.79$\pm$0.19       & 2.744     \\
   $\Omega$+$\bar\Omega$     & 0.94$\pm$0.15       & 0.9498     \\
\hline
\end{tabular}\end{center}
\caption{Comparison between calculated, at the end of hydrodynamical stage, 
and fitted midrapidity densities of particle species in most central Pb+Pb 
collisions at \snn = 2.76 TeV. The relative errors on calculated multiplicities 
are the same as the experimental measurements at the same centrality. The
$\gamma_S$ paramter has been fixed to 1. The overall normalization is arbitrary.}
\label{check}
\end{table}

The modification factors for $\pi^+$, proton, and $\Xi^-$ are shown in fig.~\ref{urqmfact} 
as a function of the collision centrality. Note that the modification factors get 
closer to 1 for peripheral collisions. 

At this point it is important to discuss the importance of multiparticle (= $N$ body
with $N>2$) reactions and their effect on the modification factors defined above. As 
has been pointed out in earlier studies \cite{Rapp:2000gy}, the $N \pi\rightarrow p 
+ \overline{p}$ (with $N$ being 4 or 5) reaction can be responsible for the regeneration 
of protons 
and antiprotons in the hadronic phase. Since the implementation of these multiparticle 
properties in a microscopic solution of the transport equation eq.(\ref{boltzmann})
is very difficult we have to make a quantitative estimate on the importance of this
back reaction. In ref.~\cite{steinheimer} we estimated that the multi-pion fusion 
process, at the investigated beam energy, should only account for less than $10\%$ 
regenaration of protons. This result agrees well with a recent study by Pan and Pratt 
\cite{pratt}, at the same beam energy and explicitely including detailed balance, 
which finds that even if a larger LCEP temperature of $170$ MeV is chosen, only about 
$20\%$ of all annihilated protons can be regenerated. Consequentely we can assume 
that neglecting the back reaction implies a small quantitative uncertainty in the 
modifications factors and does not basically alter our findings.

\section{Results}
\label{results}

The SHM, the relevant formulae for the calculation of midrapidity yields and the
fit procedure in relativistic heavy ion collisions at very high energy have been 
described in detail elsewhere \cite{becamann}. Here we just
note that at such a large energy, the rapidity distributions are wide enough to 
enable a determination of the thermodynamical parameters of the most central fireball,
as it was possible at \snn $> 100$ GeV. Furthermore, the antiparticle/particle ratios
measured at \snn = 2.76 TeV are consistent with 1 at all centralities, hence we have
set all chemical potentials to zero and the free parameters of the fit are 2 or 3:
temperature, normalization and, optionally, $\gamma_S$.
\begin{figure}[ht]
\begin{center}
\includegraphics[width=0.6\columnwidth]{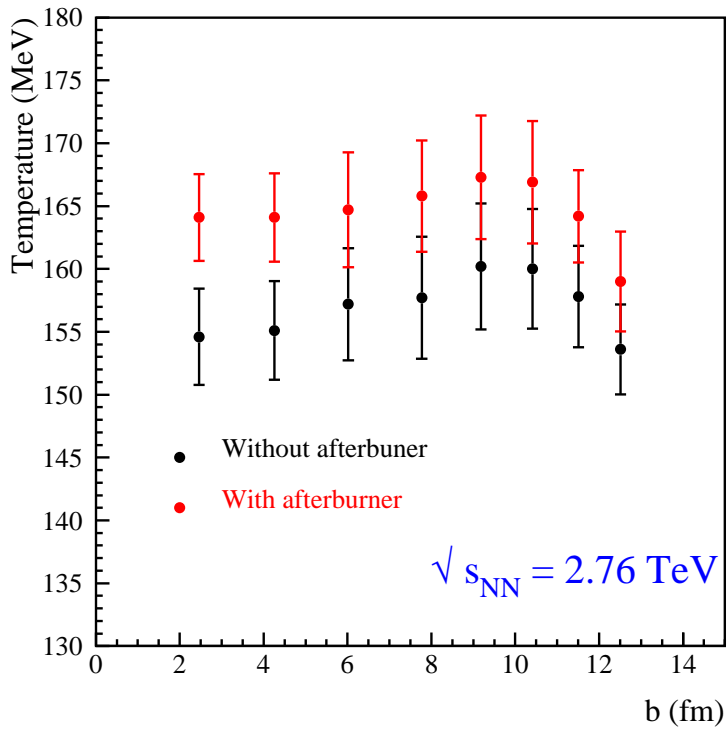}
\caption{(Color online) Temperature as a function of the impact parameter $b$ (central values
corresponding to centralities measured by ALICE). Black dots: chemical 
freeze-out temperature. Red dots: LCEP (see text) temperature obtained by including UrQMD
modification factors.}
\label{temp}
\end{center}
\end{figure} 

As a first step, we have fitted the measured multiplicities to the basic version 
of the SHM with $\gamma_S=1$ (see table~\ref{fits}). For the most central collisions, 
we confirm previous findings \cite{becafrank2} as well as recent analysis by different 
groups \cite{andronic} with a $\chi^2/dof \simeq 17/8$ and an overestimation of 
proton yields by about $2\sigma$ along with an underestimation of pion yield by 
$1.4 \sigma$ (see table~\ref{mults}) which seems to be a common feature of SHM fits 
to high energies \cite{various}. For the mid-peripheral bins, the fit quality is 
significantly worse, with a $\chi^2/dof \simeq 31/7$ and larger discrepancies for 
both pions and protons, as well as for $\Xi$. 
\begin{table}
\setlength{\tabcolsep}{10pt}
\centering
\begin{tabular}{c|c|c|c|c|c}
\hline\hline
\multirow{2}{*}{Centrality}	&		&	Plain SHM fit			&	Plain SHM fit 			&	SHM+afterburning			&	SHM+afterburning			\\ 
	&		&				&	$\gamma_S=1$			&				&	$\gamma_S=1$				\\ \hline
\multirow{3}{*}{0-5\%}	&	$T$ (MeV)	&	154.5	$\pm$	3.8	&	157.2	$\pm$	4.1	&	164.1	$\pm$	3.4	&	167.3	$\pm$	3.8	\\ 
	&	$\gamma_S$	&	1.082	$\pm$	0.059	&	1			&	1.071	$\pm$	0.043	&	1			\\ 
	&	$\chi^2$/dof	&	14.23/7			&	18.52/8			&	7.66/7			&	10.95/8			\\ \hline 
\multirow{3}{*}{5-10\%}	&	$T$ (MeV)	&	155.1	$\pm$	3.9	&	158.8	$\pm$	5.1	&	164.1	$\pm$	3.5	&	168.0	$\pm$	4.4	\\ 
	&	$\gamma_S$	&	1.116	$\pm$	0.058	&	1			&	1.086	$\pm$	0.042	&	1			\\ 
	&	$\chi^2$/dof	&	17.99/7			&	29.54/8				&	9.76/7			&	16.31/8			\\ \hline 
\multirow{3}{*}{10-20\%}	&	$T$ (MeV)	&	157.2	$\pm$	4.7	&	162.3	$\pm$	5.6	&	164.7	$\pm$	4.6	&	170.0	$\pm$	5.2	\\ 
	&	$\gamma_S$	&	1.128	$\pm$	0.066	&	1			&	1.099	$\pm$	0.055	&	1			\\ 
	&	$\chi^2$/dof	&	20.86/7			&	34.27/8				&	15.73/7			&	23.97/8			\\ \hline 
\multirow{3}{*}{20-30\%}	&	$T$ (MeV)	&	157.7	$\pm$	4.8	&	162.9	$\pm$	6.3	&	165.8	$\pm$	4.4	&	171.3	$\pm$	5.6	\\ 
	&	$\gamma_S$	&	1.141	$\pm$	0.071	&	1			&	1.111	$\pm$	0.053	&	1			\\ 
	&	$\chi^2$/dof	&	22.95/7			&	38.02/8				&	13.23/7			&	22.52/8			\\ \hline 
\multirow{3}{*}{30-40\%}	&	$T$ (MeV)	&	160.2	$\pm$	5.0	&	162.8	$\pm$	6.0	&	167.3	$\pm$	4.9	&	170.5	$\pm$	5.9	\\ 
	&	$\gamma_S$	&	1.113	$\pm$	0.073	&	1			&	1.099	$\pm$	0.060	&	1			\\ 
	&	$\chi^2$/dof	&	24.40/7			&	33.84/8				&	17.52/7			&	24.95/8			\\ \hline 
\multirow{3}{*}{40-50\%}	&	$T$ (MeV)	&	160.0	$\pm$	4.8	&	163.0	$\pm$	5.3	&	166.8	$\pm$	4.9	&	169.8	$\pm$	5.1	\\ 
	&	$\gamma_S$	&	1.088	$\pm$	0.065	&	1			&	1.068	$\pm$	0.057	&	1			\\ 
	&	$\chi^2$/dof	&	20.71/7			&	26.72/8				&	16.21/7			&	19.85/8			\\ \hline 
\multirow{3}{*}{50-60\%}	&	$T$ (MeV)	&	157.8	$\pm$	4.0	&	157.8	$\pm$	4.0	&	164.1	$\pm$	3.7	&	163.8	$\pm$	3.6	\\ 
	&	$\gamma_S$	&	0.999	$\pm$	0.058	&	1			&	0.980	$\pm$	0.045	&	1			\\ 
	&	$\chi^2$	&	12.85/7			&	12.85/8				&	8.21/7				&	8.44/8			\\ \hline 
\multirow{3}{*}{60-70\%}	&	$T$ (MeV)	&	153.6	$\pm$	3.6	&	153.3	$\pm$	5.2	&	159.0	$\pm$	4.0	&	157.8	$\pm$	5.2	\\ 
	&	$\gamma_S$	&	0.843	$\pm$	0.051	&	1			&	0.843	$\pm$	0.050	&	1			\\ 
	&	$\chi^2$/dof	&	5.59/7			&	14.36/8				&	3.13/7				&	12.39/8			\\ \hline \hline
\end{tabular}
\caption{Results of the fits to the SHM for different centralities at \snn = 2.76 TeV. 
First and second column: results of the traditional fit to SHM with $\gamma_S$ and with 
$\gamma_S$ fixed to 1. Third and fourth column: same, but with afterburning corrections
to the theoretical yields.}
\label{fits}
\end{table} 
It has been shown that in peripheral bins \cite{starfits,becamann} the corona of 
single NN collisions makes strange particle yields lower than expected from a source
at full chemical equilibrium. Therefore, we introduce $\gamma_S$ as a free parameter 
to take the effect of corona into account. This, as expected, improves the fit quality
(see table~\ref{fits}) considerably in the most peripheral bin where corona effect 
is more important. In other bins, it improves the fit, although not enough to make it
statistically significant.
\begin{table}[!ht]\begin{center}
\begin{tabular}{c|c|c|c|c|c}
\hline\hline
   \multirow{2}{*}{Particle} &   Measurement  & Plain SHM fit &   Plain SHM   &  SHM+afterburning&   SHM+afterburning \\
                             &                & $\gamma_S=1$  &               & $\gamma_S=1$     &                     \\
\hline
   $\pi^{+}$                 & 733    $\pm$ 54    &   659.2       &  645.7     & 694.2    &   683.2     \\
   $\pi^{-}$                 & 732    $\pm$ 52    &   659.2       &  645.7     & 694.2    &   683.2      \\
   K$^{+}$                   & 109.0  $\pm$  9.0  &   116.0       &  121.2     & 112.1    &   116.8      \\
   K$^{-}$                   & 109.0  $\pm$  9.0  &   116.0       &  121.2     & 112.1    &   116.8      \\
   p                         &  34.0  $\pm$  3.0  &    39.69      &   36.64    &  38.62   &   35.93       \\
   $\bar{\rm p}$             &  33.0  $\pm$  3.0  &    39.69      &   36.64    &  38.62   &   35.93       \\
   $\Lambda$                 &  26.1  $\pm$  2.8  &    21.90      &   21.55    &  22.77   &   22.39       \\
   $\Xi^{-}$                 &   3.57 $\pm$  0.27 &     3.246     &    3.427   &  3.239   &    3.384      \\
   $\Xi^{+} $                &   3.47 $\pm$  0.26 &     3.246     &    3.427   &  3.239   &    3.384      \\
   $\Omega$+$\bar\Omega$     &   1.26 $\pm$  0.22 &     1.112     &    1.237   &   1.327  &    1.444      \\ \hline
   D                         &                    &               &   0.115    &          &    0.118      \\
\hline\hline
\end{tabular}\end{center}
\caption{Comparison between measured \cite{aliceaa} and fitted midrapidity densities 
of particle species in most central Pb+Pb collisions at \snn = 2.76 TeV. In the plain
SHM fits, either with or without $\gamma_S$, there is an overestimation of proton 
and an underestimation of pion yields. The modification factors predicted by UrQMD 
improve the agreement between data and model for those particles. Also shown the predicted
midrapidity density of deuterons assuming they are formed at hadronization according to
SHM, and (for the afterburning case) that they are later suppressed with the square of
the modification factor calculated for protons.}
\label{mults}
\end{table}
\begin{figure}[!h]
\begin{center}
\includegraphics[width=0.6\columnwidth]{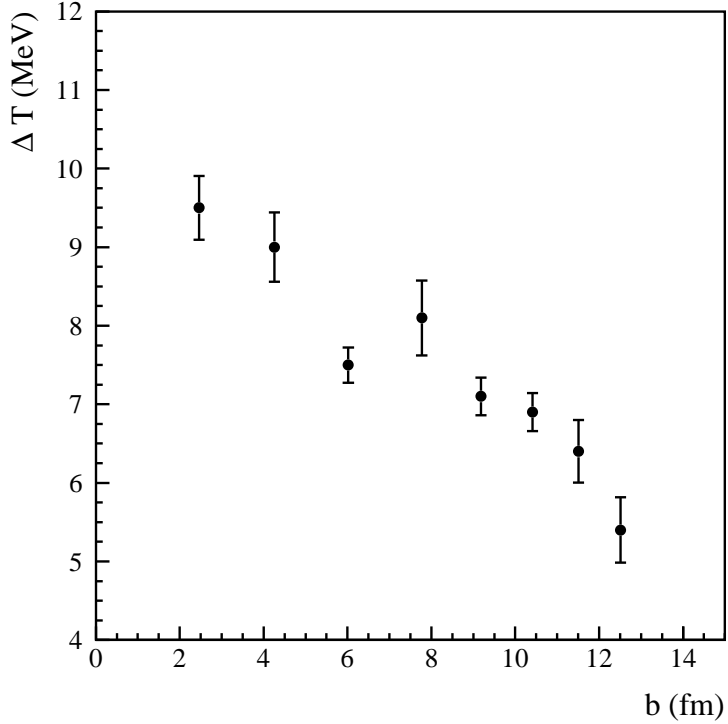}
\caption{Difference between the corrected temperature and the chemical freeze-out
temperature as a function of the impact parameter $b$ (central values corresponding 
to centralities measured by ALICE). The error bar has been estimated by taking
a 100\% correlation between the errors on $T$ in the two fits.}
\label{deltat}
\end{center}
\end{figure} 
\begin{figure}[!h]
\begin{center}
\includegraphics[width=0.6\columnwidth]{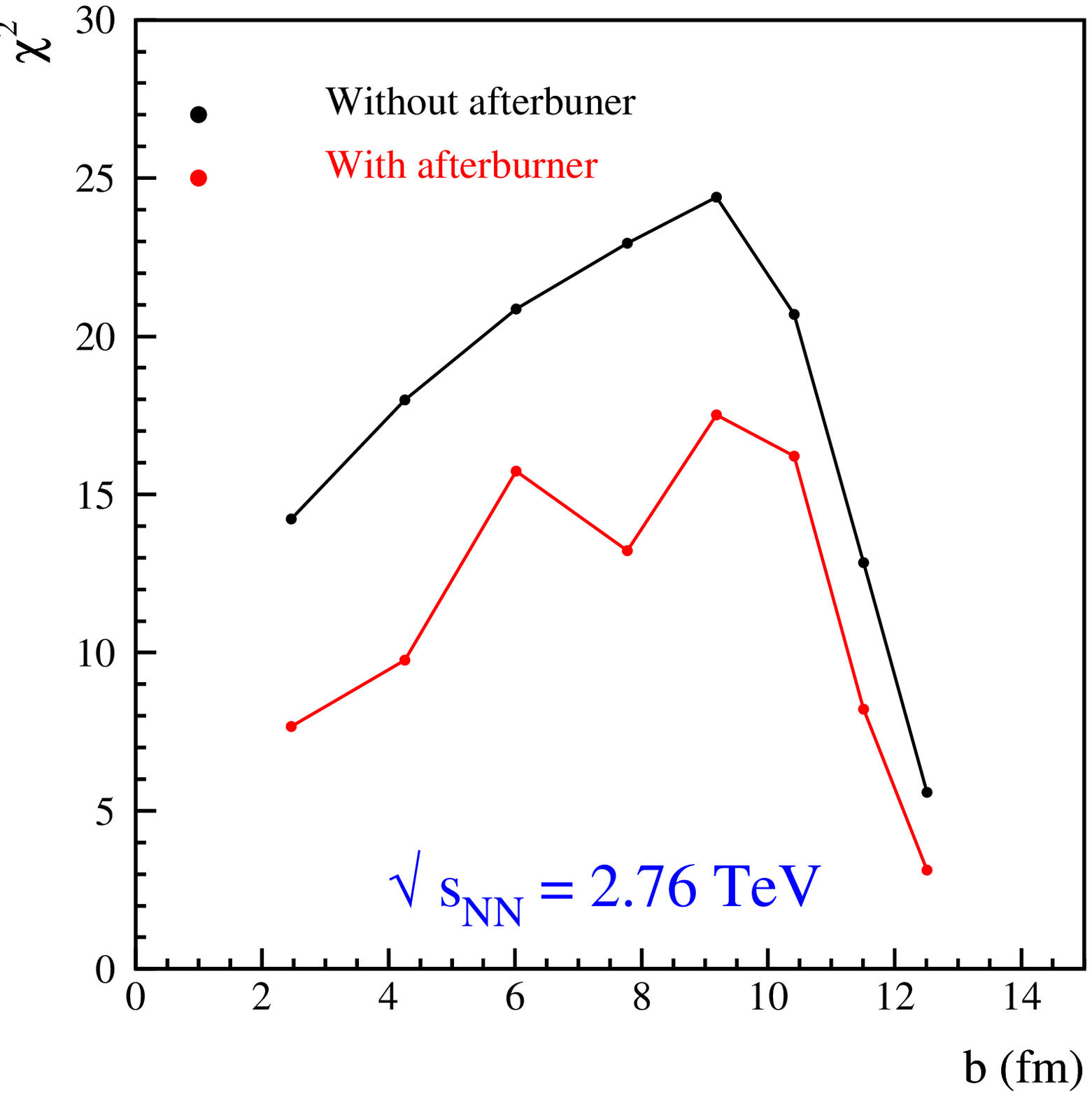}
\caption{(Color online) $\chi^2$ of the SHM fits with and without afterburning corrections as 
a function of the impact parameter $b$ (central values corresponding to centralities 
measured by ALICE). The fitted parameters being in this case $T$, $\gamma_S$
and the normalization, the number of degrees of freedom is 7.}
\label{chi2}
\end{center}
\end{figure} 

It should be noted that the chemical freeze-out temperature, in both versions with and
without $\gamma_S$, is larger in mid-peripheral bin than in central (see fig.~\ref{temp})
collisions. To assess the significance of the difference one should take into account 
that the errors on fit parameters are strongly correlated, as it is apparent from 
fig.~\ref{temp} because so are the errors on particle multiplicities measured in the
different centrality bins. The increase of temperature toward peripheral bins is observed 
for the first time and it is in qualitative agreement with the idea of an afterburning 
stage, which, 
if present, has to depend on the total multiplicity as discussed in sect.~\ref{fo}. 
Thus, the more central the collision, the longer the time spent into the colliding 
hadronic stage and the larger the shift from hadronization temperature (assumed to 
be constant) down to the chemical freeze-out. This effect was studied in a previous 
paper of ours \cite{becafrank1} at SPS energies. 

The results of the fit including corrections for afterburning are shown in the third 
column of table~\ref{fits}. The theoretical yields are calculated multiplying the 
output from SHM (after strong and electromagnetic decays) by the modification factors 
defined in the previous section. Therefore, the fitted thermodynamical parameters 
(essentially temperature) supposedly pertain to the source at its latest state of 
chemical equilibrium, i.e. LCEP, before hadronic collisions set in. 
In a more refined calculation, one would use the thus determined LCEP conditions to 
compute modification factors with isothermal Cooper-Frye transition at that temperature 
and refit the LCEP temperature until the procedure converge. Nevertheless, already 
in the present calculation, the fitted temperature at hydro-UrQMD transition (158.2 
MeV, see sect.~\ref{analysis}) is close to the final fitted value of 164 MeV, 
showing that we are not far from full self-consistency. The improved calculations 
are already in progress \cite{progress}.

As it can be see from fig.~\ref{chi2}, the fit quality improves throughout after
the implementation of afterburning corrections. The fitted temperature rises by 
several MeV's, as shown in fig.~\ref{temp}, in agreement with our previous findings 
\cite{becafrank2}. Furthermore, the LCEP temperature is less centrality dependent 
than the plain chemical freeze-out temperature, which bears out the idea of a universal 
(at fixed baryon density) hadronization temperature \cite{becareview,satz}. This 
is best seen in fig.~\ref{deltat} where we show the difference between the corrected 
temperature and the plain SHM fitted one. The difference steadily decreases towards 
peripheral collisions, again in full agreement with the picture that afterburning 
affects less the chemical composition if the overall multiplicity is lower. There 
remain two small structures in the temperature vs centrality plot after the afterburning 
correction: a mild rise towards mid-peripheral collisions (see fig.~\ref{temp}) and 
a sizeable decrease in most peripheral collisions. The former could
be a residual of the similar behaviour seen in the plain fits that the presently 
calculated correction factors were not able to completely remove. Hopefully, a new 
calculation of correction factors with isothermal hydro-UrQMD transition \cite{progress}
could be able to work it out. The latter could be, on the other hand, a spurious 
corona effect of superposition of NN collisions with hadrons from the plasma which
we are not presently able to understand in detail. Both effects will be the subject
of further investigation.

\section{Conclusions}

To summarize, we have demonstrated that in the high multiplicity environment of 
relativistic heavy ion collisions at \snn= 2.76 TeV the inelastic collisions play
a significant role in modifying the primordial hadronic yields from hadronization.
The amount of inelastic rescattering is expected to depend on multiplicity, hence
on centrality. This effect is clearly seen in the centrality dependence of specific
particle ratios measured by the ALICE experiment and especially $\Xi/\pi$ which
- for the first time - is observed to increase towards peripheral collisions before 
dropping. 
In the framework of the statistical hadronization model, this phenomenon implies 
a slight dependence of the chemical freeze-out temperature as a function of centrality,
which is actually observed. Once suitable correction factors, estimated through the 
transport model UrQMD, are introduced, primordial particle multiplicities turn out 
to be in better agreement with a chemically equilibrated source at a nearly 
constant temperature of about 164 MeV. The difference between the chemical freeze-out 
temperature and the reconstructed latest chemical equilibrium temperature, arguably
coinciding with the hadronization temperature, decreases smoothly from central to 
peripheral collisions, as expected in this picture. Further calculations
of modification factors are in preparation to investigate the remaining small structures
seen in the behaviour of temperature as a function of centrality.
These findings are in excellent agreement with the concept of a universal statistical 
hadronization occurring at the pseudo-critical QCD temperature. 

\section*{Acknowledgments}

Part of the work of F.B. was carried out over a sabbatical at the University
of Frankfurt and FIAS, Frankfurt. We acknowledge the support of the Deutsche 
Forschungsgemeinschaft (DFG), the Hessian LOEWE initiative through HIC for FAIR
and the Istituto Nazionale di Fisica Nucleare (INFN). Computational resources 
were provided by the LOEWE Frankfurt Center for Scientific Computing (LOEWE-CSC).


\end{document}